\documentstyle[11pt,fleqn]{article}
\begin{document}
\baselineskip=8mm

\noindent
{\Large \bf A Generic Dynamical Model of Gamma-ray Burst Remnants}
 
\vspace*{20mm}
\noindent
{\bf Y.F. Huang$^{1}$, Z.G. Dai$^{1}$ and T. Lu$^{1,2,3}$}
 
\noindent
{\sl $^{1}$ Department of Astronomy, Nanjing University,
Nanjing 210093, P. R. China}
 
\noindent
{\sl $^{2}$ CCAST (World Laboratory), P.O.Box.8730, Beijing 100080,
		 P. R. China}
 
\noindent
{\sl $^{3}$ LCRHEA, IHEP, CAS, Beijing 100039, P. R. China }

\vspace{20mm}

\noindent 
{\bf ABSTRACT \\ \\}
The conventional generic model is deemed to explain the dynamics of 
$\gamma$-ray burst remnants very well, no matter whether they are adiabatic 
or highly radiative. However, we find that for adiabatic expansion, the 
model could not reproduce the Sedov solution in the non-relativistic phase, 
thus the model needs to be revised. In the present paper, a new differential
equation is derived. The generic model based on this equation has been 
shown to be correct for both radiative and adiabatic fireballs, and in both 
ultra-relativistic and non-relativistic phase. 

\vspace{10mm}

\newpage

\section{\bf INTRODUCTION}
Since the BeppoSAX detection of GRB 970228, X-ray afterglows have been 
observed from about 15 gamma-ray bursts (GRBs), of which 10 were detected 
optically and 5 even also in radio wavelengths 
(Costa et al. 1997; Kulkarni et al. 1998; Bloom et al. 1998; 
Piran 1998; and references 
therein). The cosmological origin of at least some GRBs is thus firmly 
established. The so called fireball model 
(Goodman 1986; Paczy\'{n}ski 1986; Rees \& M\'{e}sz\'{a}ros 1992, 1994; 
M\'{e}sz\'{a}ros \& Rees 1992; Katz 1994; Sari, Narayan \& Piran 1996) 
is strongly 
favoured, which is found successful at explaining the major features 
of the low energy light curves 
(M\'{e}sz\'{a}ros \& Rees 1997; Vietri 1997; 
Tavani 1997; Waxman 1997; Wijers, Rees \& M\'{e}sz\'{a}ros 1997; Sari 1997;
Huang et al. 1998; Dai \& Lu 1998a; 
Dai, Huang \& Lu 1998). 
A variant of this model, where central engines (e.g., strongly magnetized 
millisecond pulsars) supply energy to postburst fireballs through magnetic 
dipole radiation, has been proposed to account for the special features 
of the optical afterglows from GRB 970228 and 
GRB 970508 (Dai \& Lu 1998b, c).

Since the expansion of a fireball may be either adiabatic or highly radiative, 
extensive attempts have been made to find a common model applicable 
for both cases (Blandford \& McKee 1976; Chiang \& Dermer 1998; Piran 1998).
As a result, a conventional model was suggested 
by various authors (see for example, Chiang \& Dermer 1998; Piran 1998). 
A dynamical model should be correct not only in the initial 
ultra-relativistic phase, which is well described by those simple scaling laws 
(M\'{e}sz\'{a}ros \& Rees 1997; Vietri 1997; Waxman 1997), 
but also in the consequent non-relativistic phase, which is correctly discussed 
by using the Sedov solution (Sedov 1969; Wijers, Rees \& M\'{e}sz\'{a}ros 1997).
Although the conventional model is correct for the ultra-relativistic phase, we find 
it could not match the Sedov solution in the non-relativistic limit.
So in this paper, we will construct a dynamical model that is really 
capable of describing generic fireballs, no matter whether they are 
radiative or adiabatic, and no matter whether they are ultra-relativistic 
or non-relativistic. 

\section{\bf CONVENTIONAL DYNAMIC MODEL} 
A differential equation has been proposed to depict the expansion of GRB 
remnants (Chiang \& Dermer 1998; Piran 1998), 
\begin{equation}
\frac{d \gamma}{d m} = - \frac{\gamma^2 - 1} {M},
\end{equation}
where $m$ is the rest mass of the swept-up medium, $\gamma$ is the bulk 
Lorentz factor and $M$ is the total mass in the co-moving frame, including 
internal energy $U$. Since thermal energy produced during the collisions 
is $ dE = (\gamma -1) dm \: c^2 $, usually we assume: 
$dM = (1 - \epsilon) dE / c^2 + dm = [( 1 - \epsilon) \gamma + \epsilon ] dm$, 
where $\epsilon$ is defined as the fraction of the shock generated thermal 
energy (in the co-moving frame) that is radiated 
(Piran 1998).
It is putative that equation (1) is correct in both ultra-relativistic and 
non-relativistic phase, for both radiative and adiabatic fireballs. However,  
after careful inspection, we find that during the non-relativistic phase of 
an adiabatic expansion, equation (1) could not give out a solution consistent 
with the Sedov results (Sedov 1969).

\subsection{\bf Radiative case}

In the highly radiative case, 
$\epsilon = 1$, $dM = dm$, equation (1) reduces to, 
\begin{equation}
\frac{d\gamma}{dm} = - \frac{\gamma^2 - 1}{M_{\rm ej} + m}, 
\end{equation}
where $M_{\rm ej}$ is the mass ejected from the GRB central engine.  
Then an analytic solution is available
(Blandford \& McKee 1976; Piran 1998):
\begin{equation}
\frac{(\gamma - 1) (\gamma_0 +1)}{(\gamma + 1) (\gamma_0 - 1)} = 
   \left ( \frac{m_0 + M_{\rm ej}}{m + M_{\rm ej}} \right )^2,
\end{equation}
where $\gamma_0$ and $m_0$ are initial values of $\gamma$ and $m$ 
respectively. Usually we assume $\gamma_0 \sim \eta / 2$, 
$m_0 \sim M_{\rm ej} / \eta$, where $\eta \equiv E_0 / (M_{\rm ej} c^2)$ 
and $E_0$ is the total energy in the initial fireball 
(Waxman 1997; Piran 1998). 

During the ultra-relativistic phase, $\gamma \gg 1$, $M_{\rm ej} \gg m$, 
equation (3) gives $(\gamma + 1) m \approx M_{\rm ej}$, or equivalently 
the familiar power-law $\gamma \propto R^{-3}$, where $R$ is the radius of 
the blast wave. In the later non-relativistic phase, 
$\gamma \sim 1$, $m \gg M_{\rm ej}$, we have:
$m^2 \beta^2 = 4 M_{\rm ej}^2$, or $\beta \propto R^{-3}$, where 
$\beta = v/c$ and $v$ is the bulk velocity of the material. This is 
consistent with the late isothermal phase of the expansion of supernova
remnants (SNRs) (Spitzer 1968). From these approximations, we believe  
that equation (2) is really correct for highly radiative 
fireballs.

\subsection{\bf Adiabatic case}

In the adiabatic case, $\epsilon = 0$, $dM = \gamma dm$, equation (1) 
also has an analytic solution (Chiang \& Dermer 1998) : 
\begin{equation}
M = [ M_{\rm ej}^2 + 2 \gamma_0 M_{\rm ej} m + m^2]^{1/2},
\end{equation}
\begin{equation}
\gamma = \frac{m + \gamma_0 M_{\rm ej}}{M}. 
\end{equation}
During the ultra-relativistic phase, 
$\gamma_0 M_{\rm ej} \gg m \gg M_{\rm ej}/ \gamma_0$,  
$\gamma \gg 1$, this solution can produce the familiar power-law 
$\gamma \propto R^{-3/2}$, which is often quoted for an adiabatic 
blastwave decelerating in a uniform medium. In the non-relativistic limit 
($\gamma \sim 1$, $m \gg \gamma_0 M_{\rm ej}$), Chiang \& 
Dermer (1998) have derived $\gamma \approx 1 + \gamma_0 M_{\rm ej} / m$, 
so that they believe it also agrees with the Sedov solution 
(Lozinskaya 1992).
However we find that their approximation is not accurate enough, because 
they have omitted some first-order infinitesimals of 
$\gamma_0 M_{\rm ej} / m$. The correct approximation could be obtained 
only by retaining all the first and second order infinitesimals, which in  
fact gives: $\gamma \approx 1 + (\gamma_0 M_{\rm ej}/m)^2 /2$, then we have   
$\beta \propto R^{-3}$. This is not consistent with the Sedov solution! 
We have also evaluated equation (1) numerically, the result is consistent 
with equations (4) and (5), all pointing to $\beta \propto R^{-3}$, not 
the relation of $\beta \propto R^{-3/2}$ as declared popularly in the 
literature (Chiang \& Dermer 1998; Piran 1998).
 
This question is serious. First, it means that equation (1) is not a 
dependable model for non-radiative fireballs, although it can 
reproduce the major features in the ultra-relativistic phase. Second, 
the expansion of a realistic fireball is widely believed to be highly 
radiative at first, but after only several days, the expansion will 
become non-radiative (Sari, Piran \& Narayan 1998; Dai, Huang \& Lu 1998). 
So in the non-relativistic phase, the fireball 
is more likely to be adiabatic rather than highly radiative. 
However, it is just in this condition that the conventional model fails. 
So any calculation made according to equation (1) will lead to serious 
deviations in the light curves in the non-relativistic phase.

\section{\bf OUR GENERIC MODEL}

Equation (1) is not consistent with the Sedov solution, we need to revise 
it. In the fixed frame, since the total kinetic energy of the fireball is 
$E_{\rm K} = (\gamma - 1) (M_{\rm ej} + m) c^2 + (1 - \epsilon) \gamma U$ 
(Panaitescu, M\'{e}sz\'{a}ros \& Rees 1998), 
and the radiated thermal energy 
is $\epsilon \gamma (\gamma - 1) dm \: c^2$ (Blandford \& McKee 1976),
we have: 
\begin{equation}
d[ (\gamma - 1)(M_{\rm ej} + m) c^2 + (1 - \epsilon) \gamma U] 
     = - \epsilon \gamma (\gamma - 1) dm \: c^2. 
\end{equation}
For the item $U$, it is usually 
assumed: $d U = (\gamma - 1) dm \: c^2$ 
(Panaitescu, M\'{e}sz\'{a}ros \& Rees 1998). 
Equation (1) has been derived just in this way. 
However, the jump conditions (Blandford \& McKee 1976) at the forward 
shock imply that $U = (\gamma - 1) m c^2 $, so we suggest that the correct 
expression for $dU$ should be: 
$dU = d[(\gamma - 1) m c^2] = (\gamma - 1) dm \: c^2 + m c^2 d \gamma$. 
Here we simply 
use $U = (\gamma - 1) m c^2$ and substitute it into equation (6), 
then it is easy to get: 
\begin{equation}
\frac{d \gamma}{d m} = - \frac{\gamma^2 - 1}
       {M_{\rm ej} + \epsilon m + 2 ( 1 - \epsilon) \gamma m}. 
\end{equation}
We expect this equation should describe a generic fireball 
correctly. 

Indeed, in the highly radiative case ($\epsilon = 1$), equation (7) 
reduces to equation (2) exactly. While in the adiabatic case 
($\epsilon = 0$), equation (7) reduces to : 
\begin{equation}
\frac{d \gamma}{d m} = - \frac{\gamma^2 - 1}
       {M_{\rm ej} + 2 \gamma m}. 
\end{equation}
This equation has an analytic solution: 
\begin{equation}
(\gamma - 1) M_{\rm ej} c^2 + (\gamma^2 - 1) m c^2 \equiv E_{\rm K0},
\end{equation}
where $E_{\rm K0}$ is the initial value of $E_{\rm K}$. In the 
ultra-relativistic 
phase ($\gamma_0 M_{\rm ej} \gg m \gg M_{\rm ej}/ \gamma$), 
we get the 
familiar relation of $\gamma \propto R^{-3/2}$. And in the 
non-relativistic phase ($m \gg M_{\rm ej}$), we 
get $\beta \propto R^{-3/2}$ as required by the Sedov 
solution. 

For any other $\epsilon$ value between 0 and 1, equation (7) describes 
the evolution of a partially radiative fireball. Unfortunately,  
we now could not find an exact analytic solution for equation (7). 
But in the non-relativistic phase, by assuming $m \gg M_{\rm ej}$, 
we still can get $m (\gamma - 1)^{(2 - \epsilon)/2} \equiv {\rm const}$, 
that is: 
\begin{equation}
\beta \propto R^{-3/(2 - \epsilon)}. 
\end{equation}

\section{\bf NUMERICAL RESULTS}

We have evaluated equation (7) numerically, bearing in mind 
that (Huang et al. 1998):
\begin{equation}
dm = 4 \pi R^2 n m_{\rm p} dR,
\end{equation}
\begin{equation}
dR = \beta c \gamma (\gamma + \sqrt{\gamma^2 -1}) dt,
\end{equation}
where $n$ is the number density of interstellar medium, $m_{\rm p}$
is the mass of proton, and $t$ is the time measured by an observer. We 
take $E_0 = 10^{52}$ ergs, $n = 1$ cm$^{-3}$, 
$M_{\rm ej} = 2 \times 10^{-5}$ M$_{\odot}$. Figures (1) -- (4) illustrate 
the evolution of $\gamma$, $v$, $R$, and $E_{\rm K}$ respectively. In 
these figures, we have set $\epsilon = 0$ (full lines), 0.5 (dotted 
lines), and 1 (dashed lines). It is clearly shown that our generic 
model overcomes the shortcoming of equation (1).

For example, for highly radiative expansion, the dashed lines in these 
figures approximately satisfy $\gamma \propto t^{-3/7}$, 
$R \propto t^{1/7}$, $\gamma \propto R^{-3}$, 
$E_{\rm K} \propto t^{-3/7}$ 
when $\gamma \gg 1$, 
and $v \propto t^{-3/4}$, $R \propto t^{1/4}$, $v \propto R^{-3}$,  
$E_{\rm K} \propto t^{-3/4}$ when $\gamma \sim 1$. While for adiabatic 
expansion, the full lines satisfy $\gamma \propto t^{-3/8}$, 
$R \propto t^{1/4}$, $\gamma \propto R^{-3/2}$ when $\gamma \gg 1$, 
and satisfy $v \propto t^{-3/5}$, $R \propto t^{2/5}$, $v \propto R^{-3/2}$ 
when $\gamma \sim 1$.

\section{\bf CONCLUSION AND DISCUSSION}

The conventional dynamic model is successful at describing highly 
radiative GRB remnants, however it has difficulty in reproducing the 
Sedov solution for adiabatic fireballs. This is completely unnoticed   
in the literature. We have constructed a new generic model to overcome 
this shortcoming. Numerical evaluation has proved that our model is highly 
credible. We hope this work would remind researchers of the importance 
of the transition from ultra-relativistic to non-relativistic phase, 
which might occur as early as $10^6$ -- $10^7$ s since the initial 
burst (Huang, Dai \& Lu 1998). 

In the above analysis, for simplicity, we have assumed that $\epsilon$ is 
a constant. But in realistic fireballs, $\epsilon$ is expected to 
evolve from 1 to 0 due to the changes in the relative importance of
synchrotron-radiation-induced and expansion-induced loss of energy 
(Dai, Huang \& Lu 1998). Assuming electrons in the co-moving frame carry 
a fraction $\xi_{\rm e} = 1$ of the total thermal energy and that the 
magnetic energy density is a fraction $\xi_{\rm B}^2 = 0.01$ of it, we 
re-evaluate equation (6) numerically. The results are plotted in 
Figs (1) -- (4) with dash-dotted lines. We see from Fig. (4) that 
the evolution of $\epsilon$ changes $E_{\rm K} (t)$ dramatically. 

It is worth mentioning that SNRs evolve from non-radiative to radiative 
stage, but GRB remnants are just on the contrary. This is not surprising 
because GRB remnants radiate mainly through synchrotron radiation while 
SNRs lose energy due to excited ions. It is reasonable to deduce that 
at very late stages, when the cooling due to ions becomes important, GRB 
remnants may become highly radiative again, just in the same way that 
SNRs do. The transition may occur when the temperature drops to below 
$\sim 10^6$ K and the velocity is just several tens kilometer per second. 
This needs to be addressed in more detail.

Another interesting problem is the possibility that HI supershells might 
be highly evolved GRB remnants 
(Loeb \& Perna 1998; Efremov, Elmegreen \& Hodge 1998). 
Our Figs (3) and (4) have shown that typical adiabatic 
GRB fireballs can evolve to $R \sim 1$ kpc at $t \sim 10^6$ -- $10^7$ 
yr, with $v \sim 10$ km/s, but highly radiative fireballs are obviously  
not powerful enough. To discuss this possibility in detail, we should
pay attention to the possible adiabatic-to-radiative 
transition mentioned just above. 

\vspace{5mm}

We are very grateful to R. Wijers for his valuable comments and suggestions.
This work was partly supported by the National Natural Science 
Foundation of China, grants 19773007 and 19825109, and the National Climbing 
Project on Fundamental Researches.

\newpage

\noindent
{\bf REFERENCES}

\noindent
\begin{description}
\item Blandford R.D., McKee C.F., 1976, Phys. Fluids, 19, 1130
\item Bloom J.S. et al., 1998, ApJ, submitted (astro-ph/9808319)
\item Chiang J., Dermer, C.D., 1998, ApJ, submitted (astro-ph/9803339)
\item Dai Z.G., Huang Y.F., Lu, T., 1998, ApJ, in press (astro-ph/9806334)
\item Dai Z.G., Lu T., 1998a, MNRAS, 298, 87 
\item Dai Z.G., Lu T., 1998b, A\&A, 333, L87 
\item Dai Z.G., Lu T., 1998c, Phys Rev Lett, 81, 4301 
\item Costa E. et al., 1997, Nat, 387, 783
\item Efremov Y.N., Elmegreen B.G., Hodge P.W., 1998, 
      ApJ, in press (astro-ph/9805236)
\item Goodman J., 1986, ApJ, 308, L47 
\item Huang Y.F., Dai Z.G., Lu T., 1998, A\&A, 336, L69
\item Huang Y.F., Dai Z.G., Wei D.M., Lu T., 1998, MNRAS, 298, 459
\item Katz J., 1994, ApJ, 422, 248 
\item Kulkarni S.R. et al., 1998, Nat, 393, 35 
\item Loeb A., Perna R., 1998, ApJ, in press (astro-ph/9805139) 
\item Lozinskaya T.A., 1992, Supernovae and 
  Stellar Winds in the Interstellar Medium (New York, AIP), Chap. 9
\item M\'{e}sz\'{a}ros P., Rees M.J., 1992, MNRAS, 257, 29P 
\item M\'{e}sz\'{a}ros P., Rees M.J., 1997, ApJ, 476, 232
\item Paczy\'{n}ski B., 1986, ApJ, 308, L43 
\item Panaitescu A., M\'{e}sz\'{a}ros P., Rees M.J., 1998, 
      ApJ, in press (astro-ph/9801258) 
\item Piran T., 1998, Phys Report, in press (astro-ph/9810256)
\item Rees M.J., M\'{e}sz\'{a}ros P., 1992, MNRAS, 258, 41P 
\item Rees M.J., M\'{e}sz\'{a}ros P., 1994, ApJ, 430, L93  
\item Sari R., 1997, ApJ, 489, L37
\item Sari R., Narayan R., Piran T., 1996, ApJ, 473, 204 
\item Sari R., Piran T., Narayan R., 1998, ApJ, 497, L17
\item Sedov L., 1969, Similarity and Dimensional 
  Methods in Mechanics (Academic, New York), Chap. IV
\item Spitzer L., 1968, Diffuse Matter in Space 
  (Wiley, New York), p200 
\item Tavani M., 1997, ApJ, 483, L87
\item Vietri M., 1997, ApJ, 488, L105
\item Waxman E., 1997, ApJ, 485, L5
\item Wijers R.A.M.J., Rees M.J., M\'{e}sz\'{a}ros P., 1997, MNRAS, 288, L51 
\end{description}

\newpage 
\baselineskip=8mm

\noindent
{\large \bf Figure Captions}

\vspace{10mm}

\noindent
{\bf Figure 1.} Evolution of the bulk Lorentz factor $\gamma$. We take 
$E_0 = 10^{52}$ ergs, $n=1$ cm$^{-3}$, $M_{\rm ej} = 2 \times 10^{-5}$ 
M$_{\odot}$. The full, dotted, and dashed lines correspond to 
$\epsilon = 0$ (adiabatic), 0.5 (partially radiative), and 1 (highly 
radiative) respectively. The dash-dotted line is plotted by 
allowing $\epsilon$ to evolve with time (see Sect. (5) in the text). 

\vspace{0.5cm}
\noindent
{\bf Figure 2.} Evolution of the bulk velocity $v$. Parameters and line
styles are the same as in Fig.~1.

\vspace{0.5cm}
\noindent
{\bf Figure 3.} Evolution of the shock radius $R$. Parameters and line
styles are the same as in Fig.~1.

\vspace{0.5cm}
\noindent
{\bf Figure 4.} Evolution of the total kinetic energy $E_{\rm K}$. 
Parameters and line styles are the same as in Fig.~1.

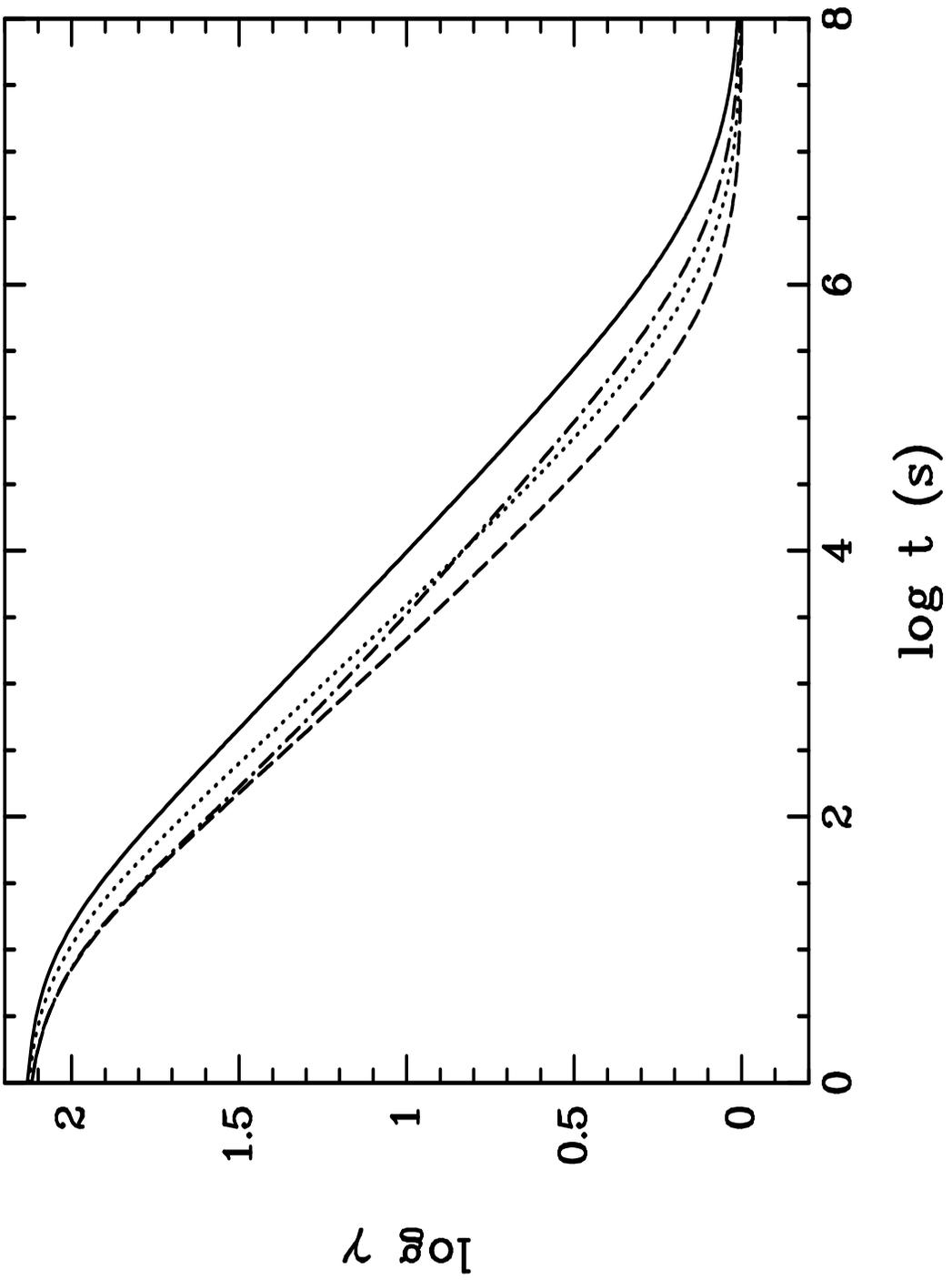
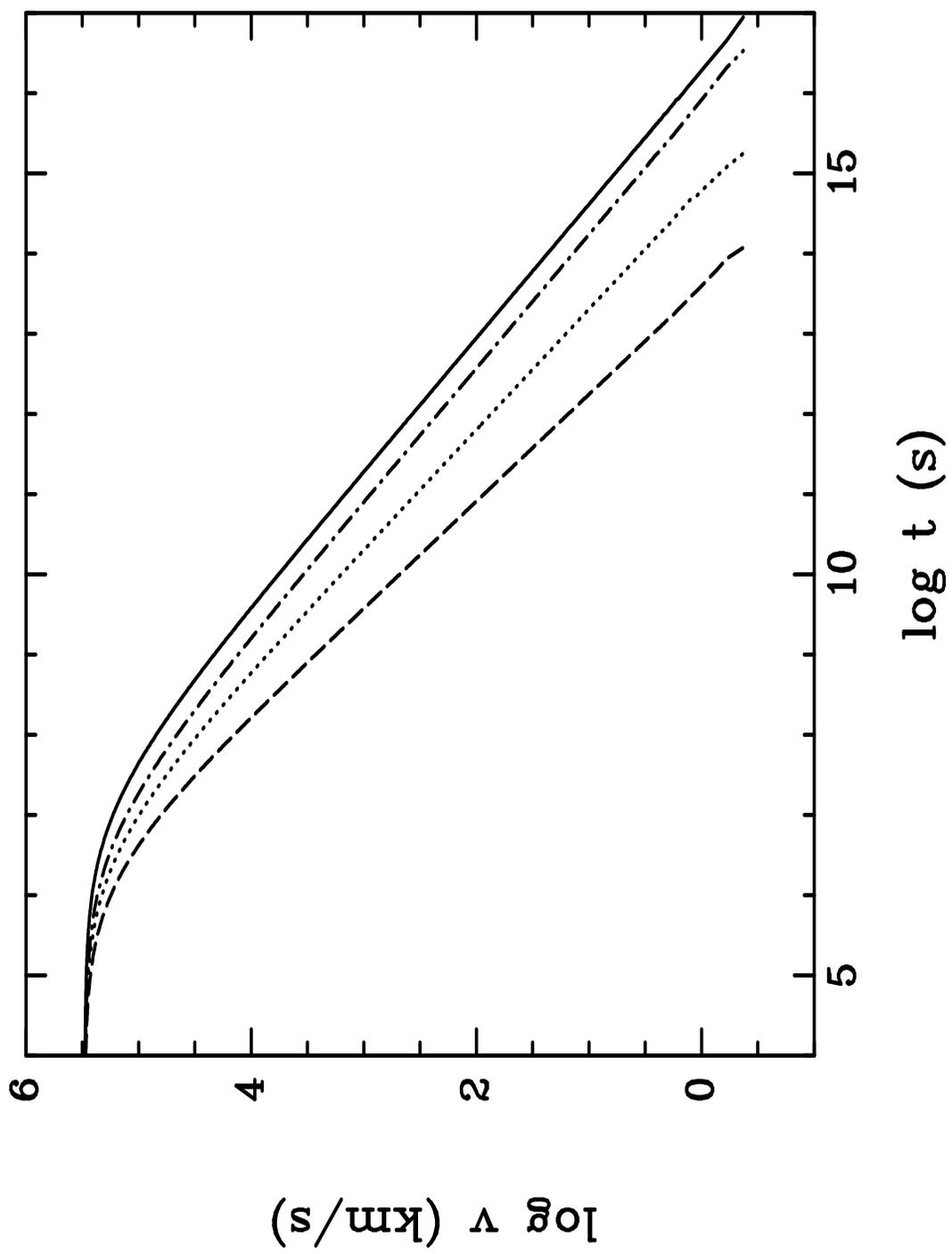
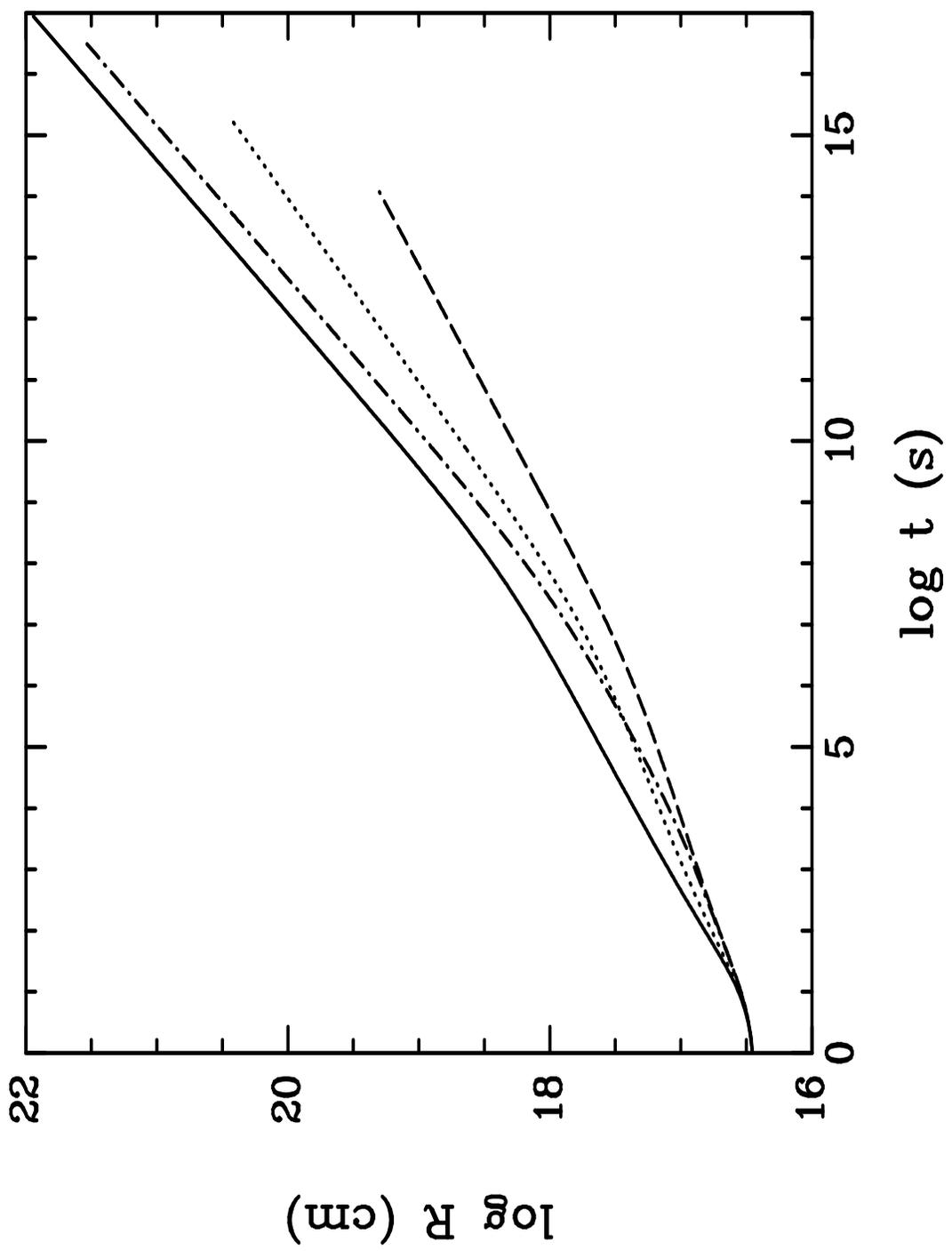
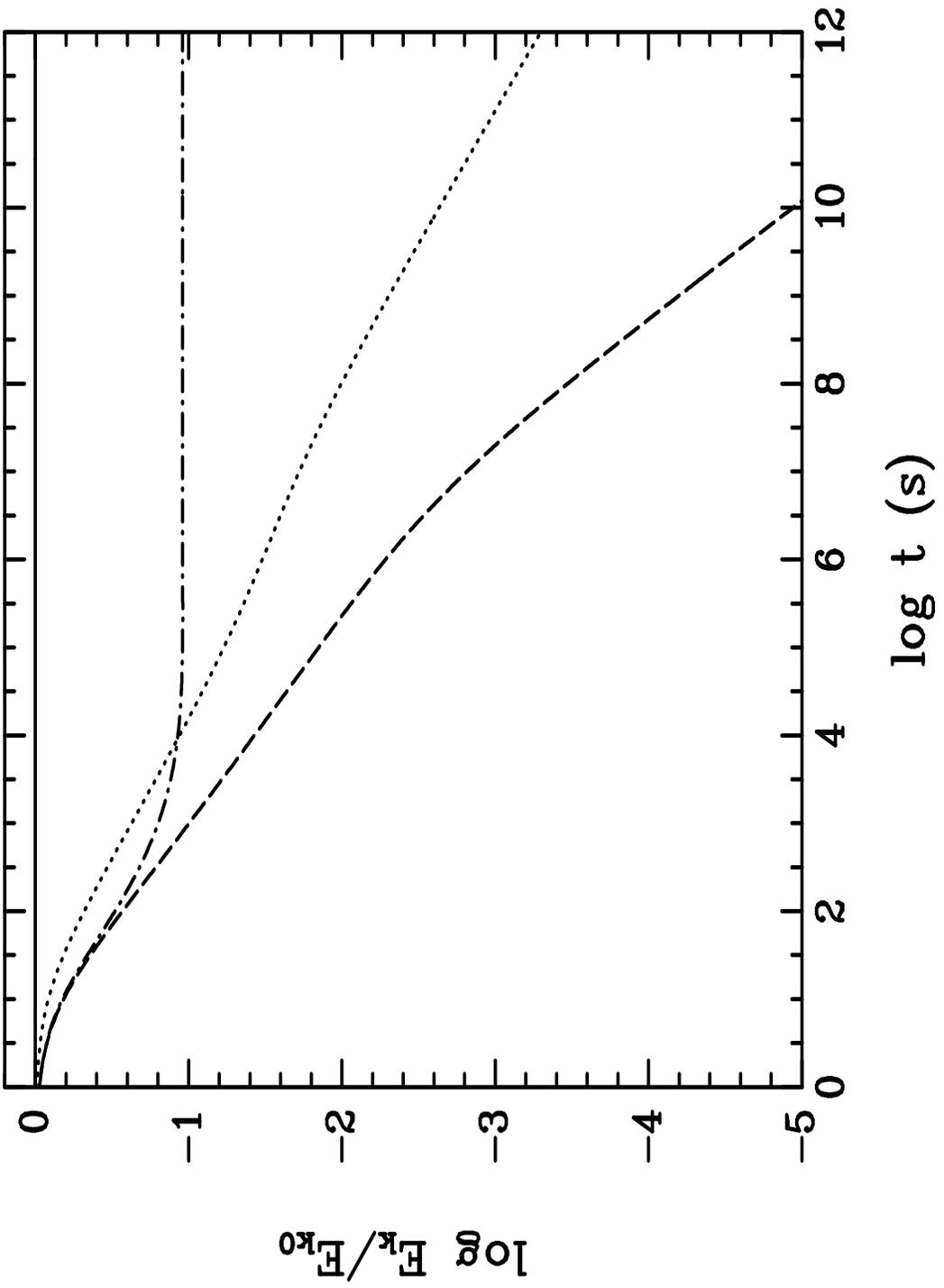
\end{document}